\documentclass[fleqn,twoside]{article}
\usepackage{espcrc2}
\usepackage{graphicx}

\title{Exploring arrival directions of UHECRs with the Yakutsk array}
\author{A.A. Ivanov\\
\vspace{1pc}
Shafer Institute for cosmophysical research \& aeronomy \\
Lenin ave., 31, Yakutsk 677980, Russia\\
e-mail: ivanov@ikfia.ysn.ru}

\begin{document}

\begin{abstract}
The data on arrival directions of ultra-high energy cosmic rays (UHECRs) detected with the Yakutsk array are analyzed. The work is induced by the recent claim of the Pierre Auger collaboration for the significant correlation found between UHECRs and positions of nearby Active Galactic Nuclei (AGN) on the celestial sphere; and no correlation the HiRes collaboration stands for. Conflicting data of four giant arrays concern possible extragalactic sources of UHECRs and appeal to the profound analysis and to the future data from the Telescope Array/Northern Auger Observatory.
\vspace{1pc}
\end{abstract}

\maketitle

\section{Introduction}
Detection of extensive air showers (EASs) of cosmic rays and the evaluation of energies, masses and arrival directions of primary particles is a crucial technique to search for UHECR sources. The localization of the sources is complicated due to charged particle trajectories distorted in unknown (extra)galactic magnetic fields. Only at the highest energies a deflection of protons, the most probable particles of cosmic rays, is less than or comparable to the angular resolution of the giant arrays.

The Pierre Auger Observatory (PAO) collaboration has recently analyzed a sample consisting of 81 EASs with energies above $40$ EeV ($=4\times10^{19}$ eV) detected from January 1, 2004 to August 31, 2007~\cite{Auger}. The authors used a part of the data (to May 27, 2006) in order to determine the parameters resulting in the maximum correlation of UHECR arrival directions with AGN. Then, the second part of the data was used to confirm the hypothesis obtained.

As a result, the observed UHECR arrival directions are found to be anisotropic, there is a significant correlation of EASs at energies above $56$ EeV within an angle of $\psi=3.1^0$ with AGN from catalog~\cite{VCV} located at distances $z\leq 0.018$\footnote{75 Mpc, assuming $H=71$\,km\,s$^{-1}$Mpc$^{-1}$.} from the Earth. In the second part of the data (from May 27, 2006), 8 of 13 EASs correlate with AGN under the same conditions that have been found for the first part, while the number of expected coincidences is 2.7 in the isotropic case, with the chance probability $P=1.7\times10^{-3}$.

This result is confirmed by the Yakutsk array data~\cite{Yak}, while the HiRes data demonstrate no significant correlation with AGN~\cite{HiRes}.

Gorbunov et al. stated contrary to the PAO hypothesis: the conclusion that the bulk of UHECRs are protons originating in nearby AGN can be rejected at 99\% CL. Instead, they attribute PAO observational data to the existence of a bright source in the direction of the Centaurus Supercluster~\cite{CenA}. Another interpretation of the data was proposed by Wibig and Wolfendale, namely, that cosmic rays are nuclei with $\overline{\ln A}=2.2.\pm0.8$ generated in radio galaxies~\cite{WW}.

In this article the situation is re-visited using available data of the giant surface arrays\footnote{e.g. PAO is a surface array where the energy is estimated via the fluorescence light; at the Yakutsk array the energy estimation is based on the Cherenkov light measurements.} and taking into account the HiRes result.

\section{The Yakutsk array and experimental data used in analysis}
The Yakutsk array geographical coordinates are $61.7^0N,129.4^0E$, 100 m above sea level\footnote{1020 g/cm$^2$.}. At present it consists of 58 ground-based and 6 underground scintillation detector stations to measure charged particles (electrons and muons) and 48 detectors of the air Cherenkov light. During more than 30 years of lifetime the array has been re-configured several times, the total area covered by detectors was maximal about 1990 ($S\sim17$ km$^2$), now it is $S\sim10$ km$^2$. During the whole observation period approximately $10^6$ showers of the primary energy above $10^{15}$ eV have been detected.

In this work a sample of the data published by Pravdin et al.~\cite{Pune} ($51$ EASs with energies above $40$ EeV and zenith angles below $60^0$) is analyzed. The angular resolution error is less than $5^0$ for these showers. The energy estimation method is based on the total flux measurement of the air Cherenkov light and the number of electrons and muons at observation level~\cite{CRIS,JETP}. The energy estimation error is about $30\%$ and  $50\%$ for the showers with axes inside the array area and in the effective region outside, respectively. In some cases a data sample was extended down to $E_{thr}=1$ EeV in order to trace the energy dependence.

\begin{figure}[t]\begin{center}
\includegraphics[width=\columnwidth]{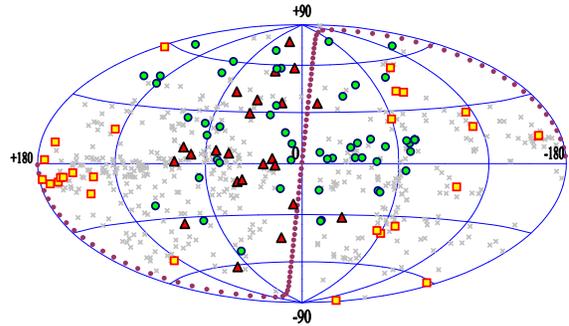}
\caption{\label{fig:map} UHECR arrival directions in supergalactic coordinates. Observational data: AGASA (circles); PAO (squares); the Yakutsk array (triangles); AGN from Veron's catalog (crosses). Galactic plane is indicated by dots.}
\end{center}\end{figure}

Additionally, the EAS events detected with AGASA (58 EASs, $E\geq40$ EeV, \cite{Uchihori}) and PAO (27 EASs, $E\geq57$ EeV, \cite{PAO}) are used in analysis. Arrival directions of UHECRs under consideration are illustrated in Fig.~\ref{fig:map} in supergalactic coordinates using Hammer-Aitoff projection, together with celestial positions of AGN from the catalog~\cite{VCV}, with redshifts $z<0.018$.

\section{Exposure of sky zones to the surface arrays}
Due to the array acceptance area depending on zenith angle and energy because of the shower attenuation in the atmosphere as wel as the geometric factor, the calculation of a celestial region exposure to the surface array in the diurnal cycle is a non-trivial task.

A simple method to calculate the exposure is the Monte Carlo algorithm implementation~\cite{Wavelet}. In this reversal approach one can take $N$ random points distributed isotropically\footnote{$f(\theta,\phi)\propto \sin2\theta$ at energies above $10$ EeV.} in the horizontal system, transform to equatorial angles and than count the points hit within the given celestial region. In the limit $N\rightarrow\infty$ the exact exposure of the region is evaluated.

\begin{figure}
\includegraphics[width=\columnwidth]{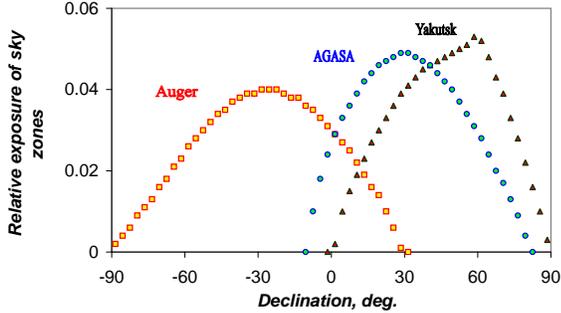}
\caption{Sky coverage by the surface arrays: AGASA ($35.8^0 N, \theta_{max}=45^0$); PAO ($32.2^0 S, \theta_{max}=60^0$) and Yakutsk ($61.7^0 N, \theta_{max}=60^0$). $E>10$ EeV.} \label{fig:Expo}
\end{figure}

In Fig.~\ref{fig:Expo} the resultant relative exposures of the right ascension zones ($\alpha\in(0^0,360^0),3^0j<\delta\leq3^0(j+1), j=-30,..,30$) to three surface arrays are shown; $N=10^7$. Fields of view are complementary in the case of the Yakutsk array and PAO, while AGASA is observing a region in low latitudes where is a depression in the consolidated fields of other two arrays.

\section{Searching for correlation with the large scale structures}
The distribution of luminous matter is highly anisotropic in the universe, notably, there are two planar structures - Milky Way disk within the Galaxy and the so-called supergalactic plane (SGP)~\cite{SGP} in the galaxy distribution. There may be a correlation between these structures and UHECR arrival directions, if the concentration of potential sources (e.g. SNRs, AGN) is proportional to the density of matter.

In this section, the Yakutsk array data are re-analyzed in comparison with the data from other arrays in order to verify this possibility. We have used the (super)galactic plane enhancement parameter, $R$, which is given by
\begin{equation}
R=\frac{n(|b|<d)-n(|b|\geq d)}{n(|b|<d)+n(|b|\geq d)},
\label{eq:GPE}\end{equation}
where $b$ is the (super)galactic latitude; $d$ is the plane border. A statistical error of the parameter is
\begin{equation}
\delta R=2\sqrt{\frac{n(|b|<d)n(|b|\geq d)}{(n(|b|<d)+n(|b|\geq d))^3}}.
\label{eq:dGPE}\end{equation}

\begin{figure}[t]
\includegraphics[width=0.9\columnwidth]{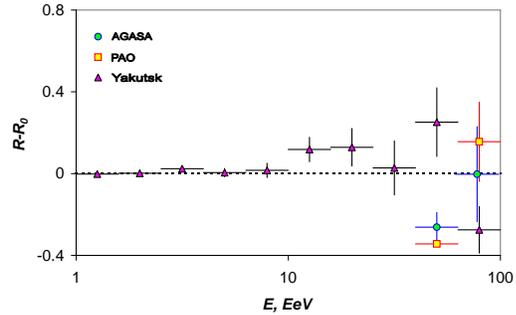}
\caption{Galactic plane enhancement difference between observed, $R$, and expected in the isotropic case, $R_0$, values. $d=10^0$. Vertical bars are statistical errors given by Eq.~\ref{eq:dGPE}, horizontal bars indicate energy bins.}
\label{fig:GPE}\end{figure}

A galactic plane enhancement (GPE) parameter is estimated using the available data of three giant arrays: AGASA, PAO and Yakutsk. In the latter case EAS events are selected with energies above 1 EeV to reveal the energy dependence of the excess flux, if any. The results are shown in Fig.~\ref{fig:GPE}. An expected parameter is calculated by the Monte Carlo method ($N=10^6$) using the algorithm described in the previous section.

There is no excess flux from the galactic disk in the whole energy range examined. The local excess in the interval $(10,20)$ EeV is insignificant. Angular dimension of the possible sources disk is uncertain, so we have scanned the range $d\in(0^0,30^0)$. None of $d$ values resulted in the significant GPE.

\begin{figure}[t]
\includegraphics[width=\columnwidth]{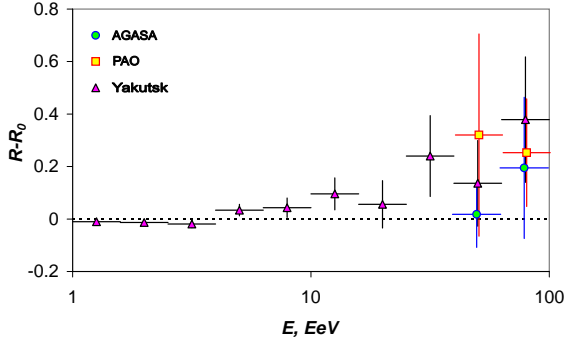}
\caption{Supergalactic plane enhancement parameter. $d=10^0$.}
\label{fig:SGPE}\end{figure}

Another possibility for the anisotropy in UHECR arrival directions is connected with the SGP, as was hypothesized by Stanev et al.~\cite{Stanev}. In order to reveal the possible excess we have calculated the enhancement parameter, SGPE, and its statistical error repeating the procedure above with definitions (\ref{eq:GPE},\ref{eq:dGPE}) for a supergalactic latitude.

A scan over the SGP angular dimension $d\in(0^0,30^0)$ revealed a maximum enhancement effect, SGPE, at $d=10^0$.
The difference with regard to the 'isotropic' value expected for the arrays is shown in Fig.~\ref{fig:SGPE} as a function of energy.

Again, there is no statistically significant enhancement from SGP, in all the data in any energy bin. But it seems to be a systematic increase of the SGPE in the Yakutsk array data which is supported by the data from AGASA and PAO above $40$ EeV: all the observational enhancement parameters are greater than those expected in the isotropic case. The possible utility of this result consists in expected evidence of the excess flux at the highest energies from future arrays, or increased number of EAS events from the PAO array.

\section{Searching for possible extragalactic sources of UHECRs}
In this section we are going to verify the PAO claim with independent datasets provided by AGASA and the Yakutsk arrays. In contrast with the previous analysis~\cite{Yak}, we have used as the null hypothesis the maximum correlation parameters given by the whole PAO dataset scanned above 40 EeV, rather than predictor-confirmer scheme dividing a sample into two parts. These parameters are: a correlation within angle $\psi=3.2^0$ between AGN at $z<0.017$ and 20 UHECRs out of the total 27 with energies above $E_{thr}=57$ EeV~\cite{PAO}.

Observed and expected-for-isotropy numbers of UHECRs correlated with AGN are calculated for the Yakutsk and AGASA EAS events applying the sky exposure (Fig.~\ref{fig:Expo}) to arrays. Resultant numbers and chance probabilities are given in Table~\ref{Table:AGN} together with the PAO result.

\begin{table}[t]
\caption{The number of coincidences, $N_{hit}$, in arrival directions of $N$ EASs above $E_{thr}$ with AGN. $N_{iso}$ and $P$ are an expected number and a chance probability for the isotropic distribution.}
\begin{center}
\begin{tabular}{lllll}
\hline
Array   & $N_{hit}$ & $N$ & $N_{iso}$ & $P,\%$ \\
\hline
PAO     & 20        & 27  & 5.6       & $4.6\times10^{-7}$ \\
AGASA   &  4        & 23  & 5.5       & 83.1 \\
Yakutsk & 12        & 24  & 4.9       & 0.12 \\ \hline
\end{tabular}\label{Table:AGN}\end{center}
\end{table}

The null hypothesis is rejected in AGASA case, while the Yakutsk array data confirm the result of the PAO collaboration, but at a lower significance level. An analysis of the HiRes data has resulted in no significant correlation with AGN~\cite{HiRes}, so the pro and con arguments are in equality. Apparently, we have to wait for the future data in this case, too.

Due to the different fields of view, as well as the energy/arrival angles estimation procedure, the 'optimal' correlation parameters can be different for the arrays. For this reason, a scan in the energy ($E>40$ EeV), redshift ($0.001<z<0.03$) and angular distance ($1^0<\psi<6^0$) of the Yakutsk array data is performed to determine the maximum ratio of the difference in the observed number of coincidences and the number expected for the isotropic case to the standard deviation. The maximum ratio appears to be reached for $22$ EASs with energies above $60$ EeV, $12$ of which arrive within $\psi=3^0$ of the AGN (while an expected number is $4.1$) at the distance from the Earth less than $z=0.015$\footnote{$63$ Mpc}. The chance probability is $P=2\times10^{-4}$. However, it is necessary to apply a penalty factor to the probability due to a posteriori selection of the parameters in this case.

\begin{figure}[t]
\center{\includegraphics[width=0.97\columnwidth]{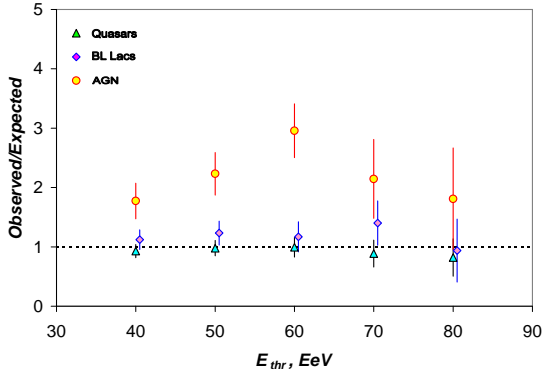}}
\caption{Ratio of the coincidences of UHECR arrival directions with extragalactic objects to the number of random coincidences expected for the isotropic distribution versus the threshold energy of the particles, $E>E_{thr}$; the Yakutsk array data. The statistical error bars are also shown.}
\label{fig:others}\end{figure}

Figure \ref{fig:others} shows the ratio of the observed number of coincidences in UHECR arrival directions (within $\psi=3^0$) with quasars, Lacertae, and AGN from the catalog~\cite{VCV} to the number of random coincidences expected for the isotropic distribution. The active nuclei are chosen at the distances $z<0.015$, the quasars are taken with redshifts $z<0.3$, and BL Lacs are selected with the luminosities $m<18$, as in~\cite{TT}. It is found that there is no significant deviation of the coincidences from isotropic expectation in the case of Lacertae and quasars. A variation in the redshift for quasars does not reveal any significant deviation. No significant correlation is also found for HP objects and BL+HP objects from the catalog~\cite{VCV}. The significant correlation with AGN is shown in Figure as was found scanning parameters above.

\begin{figure}[t]
\center{\includegraphics[width=\columnwidth]{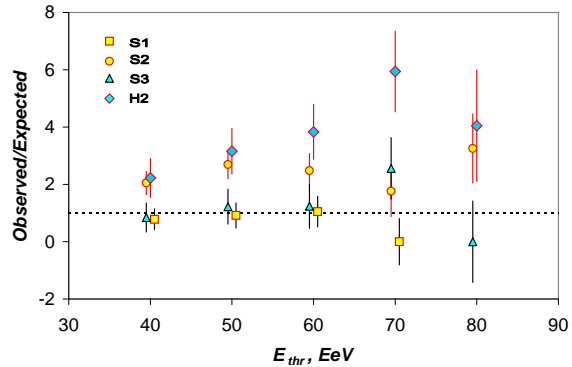}}
\caption{Ratio of UHECR correlation with different classes of AGN to the expected number.}
\label{fig:Seyferts}\end{figure}

It is interesting to know whether all classes of the objects belonging to AGN are sources or, for instance, only Seifert galaxies generate UHECRs as was suggested by Uryson~\cite{Uryson}. To answer this question, one can analyze a correlation of UHECR arrival directions with AGN classes separately. Figure~\ref{fig:Seyferts} shows the results. AGN objects are divided into four classes along the proposal of V\'{e}ron­Cetty and V\'{e}ron~\cite{VCV}: i) S1, Seifert galaxies of the first type with broad Balmer lines; ii) S2, Seifert galaxies of the second type; iii) S3, so-called LINERs (low-ionization nuclear emission-line regions), which are galaxies with weak nuclear emission lines; and iv) H2, galaxies whose spectrum of nuclear emission lines is similar to that of nebulae ionized by hot stars. The redshift boundaries giving the maximum correlation with the Yakutsk array data are selected for each class of the objects: $z<0.015$ for S1 and S3, $z<0.016$ for S2, and $z<0.024$ for H2.

Correlations of Seifert galaxies of the first type and LINERs with UHECRs do not exceed 'isotropic' expectation. Only S2 and H2 objects correlate with UHECRs (with the maxima at $E>50$ and $E>70$ EeV, respectively). The excess in the observed number of coincidences over the number expected for the isotropic distribution is $4.7\sigma$ and $3.7\sigma$ in the maxima for S2 and H2. Therefore, possible sources of UHECRs are Seifert galaxies of the second type and/or H2 objects at the distances less than $100$ Mpc.

\begin{figure}
\center{\includegraphics[width=0.9\columnwidth]{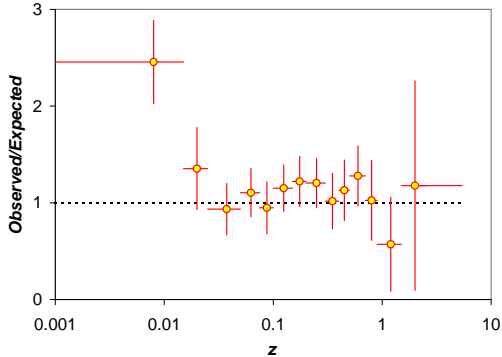}}
\caption{Ratio of the coincidences of UHECR arrival directions with AGN to the expected number in various redshift bins, $z$. The vertical bars are statistical errors. The boundaries of $z$ bins (shown by the horizontal bars) are 0.001, 0.015, 0.025, 0.05, 0.075, 0.1, 0.15, 0.2, 0.3, 0.4, 0.5, 0.7, 0.9, 1.5, and 5.4.}
\label{fig:z}\end{figure}

\section{Correlations in the redshift intervals}
Another part of the PAO hypothesis is that UHECRs correlate with AGN at distances $z<0.018$ due to the Greisen-Zatsepin-Kuzmin effect~\cite{GZK}, which strongly suppress the flux of cosmic rays with energies $E>60$ EeV from cosmological distances. In order to verify this effect with the Yakutsk array data, the ratio of the observed number of coincidences of cosmic rays ($E>60$ EeV) with AGN in various redshift intervals to the number of random coincidences expected for the isotropic distribution is used (Fig.~\ref{fig:z}). Indeed, a significant correlation of UHECRs with AGN is found only in the nearest bin $z\in(0.001,0.015)$. In all other redshift bins, the observed number of coincidences is equal (within errors) to the number expected in the isotropic case. Hence, this can be considered as one of the independent evidences of the Greisen-Zatsepin-Kuzmin effect.

\section{Corrections to the measured UHECR intensity and energy}
Correlations with AGN are found in the PAO and Yakutsk array data almost within the same angular spots ($3.2^0$ and $3^0$, respectively). At the same time, an approximate coincidence in the threshold energies at which the maximum correlation is observed by the two arrays is accidental. A comparison of the energy spectra measured with different arrays infer the existence of systematic differences between cosmic ray energies estimated. Additionally, another correction factor to the measured intensity of UHECRs is caused by the instrumental errors and fluctuations of the shower parameters\footnote{for a given primary particle energy}. The latter factor, which is derived to be~\cite{NJP}:
\begin{equation}
R_J=exp(\frac{\sigma^2\kappa^2}{2}),
\label{Eq:RJ}\end{equation}
where $\sigma$ is RMS deviation; $\kappa$ is the integral energy spectrum index, should be applied before any comparison of the energy spectra observed at different EAS arrays.

While the true energy correction factors are unknown for arrays, we can cross calibrate the energy estimation methods adjusting arbitrary factors, $R_E$, for the pairs of spectra measured, which converge them properly. The resulting spread of factors elucidates a confidence interval for the UHECR energy estimated.

\begin{table}[t]
\caption{Correction factors to energy scales of the EAS array pairs, $R_E$, averaged in the region $E>1$ EeV.}
\begin{center}
\begin{tabular}{llllll}&&&&\\
\hline
        & AGASA & HiRes & PAO & Yakutsk \\
\hline
AGASA    &    1 & 0.75 & 0.63 & 1.05 \\
HiRes    & 1.33 &    1 & 0.85 & 1.40 \\
PAO      & 1.6  & 1.2  &    1 & 1.70 \\
Yakutsk & 0.91 & 0.71 & 0.6  &    1 \\
\hline
\end{tabular}
\end{center}
\label{Table:iCalibr}
\end{table}

Table~\ref{Table:iCalibr} demonstrates the variety of correction factors to the energy estimation methods used.

\begin{figure}[t]
\center{\includegraphics[width=\columnwidth]{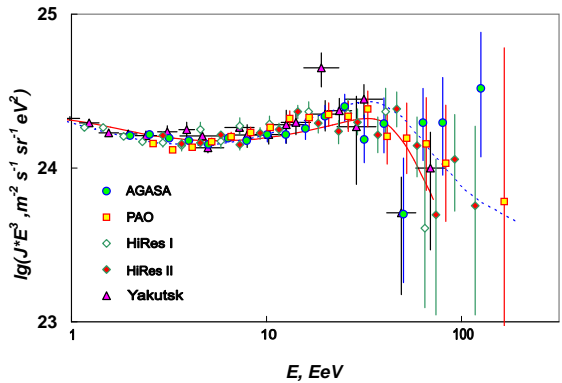}}
\caption{UHECR energy spectrum measured with giant arrays, after applying the intensity and energy corrections. The results of the extragalactic propagation models are shown by curves (details in the text).}
\label{Fig:SpectraFit}
\end{figure}

To illustrate the result of corrections $R_j\times R_E$ applied to estimated energies and intensities, the measured spectra are shown in Fig.~\ref{Fig:SpectraFit}. Energy scale factors are used here from the second column of Table~\ref{Table:iCalibr}, although any other column may be used as well.

\begin{table}[b]
\caption{Energy thresholds where the maximum correlation of UHECRs and AGN is found.}
\begin{center}
\begin{tabular}{lll}\\
\hline
        & PAO scale & Yakutsk scale \\
\hline
$E_{thr}^{PAO}$, EeV    &  57 & 97 \\
$E_{thr}^{Yak}$, EeV    &  36 & 60 \\
\hline
\end{tabular}
\end{center}
\label{Table:Threshols}
\end{table}

The UHECR propagation model results are presented by the two examples: Bahcall and Waxman model (shown by dots)~\cite{Bahcall}, and Berezinsky et al's (solid curve)~\cite{Dip07}.

In this context, the actual energy thresholds of the AGN correlation effect in the case of the Yakutsk array and PAO data seem badly different (Table~\ref{Table:Threshols}).

\section{Conclusions}
Bearing in mind negative results of AGN correlation search by AGASA and HiRes arrays; and actual difference in energy thresholds of the max correlation in the case of PAO and Yakutsk data, we can infer the preliminary conclusions:

i) AGN hypothesis of the PAO collaboration is confirmed by the Yakutsk array data;

ii) Single source models (like Cen A alone) are disfavored because PAO and Yakutsk arrays observe opposite hemispheres;

iii) S2 and H2 sub-classes of AGN are probable sources of UHECRs; other extragalactic objects exhibit no significant correlation with the Yakutsk array data;

iv) Visible UHECR sources are enclosed within ~100 Mpc area due to GZK effect.

\section*{Acknowledgements}
I would like to acknowledge the Yakutsk array staff for the data analysis and valuable discussions. The work is supported by the Russian Foundation for Basic Research (grant no. 06-02-16973).

\end{document}